\documentclass[11pt]{article}
 
\usepackage{hyperref}
\usepackage{amsmath,amsfonts,amssymb}
\hypersetup{dvips,dvipdfm,colorlinks=true,urlcolor=magenta,filecolor=magenta,linktoc=page,citecolor=red,linkcolor=blue,bookmarks=true}
\usepackage{array}
\usepackage{graphicx,epstopdf}
\usepackage{multirow}
\usepackage{xcolor}
\begin{document}
\title{\textbf{Wormholes: Myth or Reality? Prospects for Future Observations}}
\author{Subenoy Chakraborty ${ }^{1}$ \footnote{\url{schakraborty.math@gmail.com} (corresponding author)} ~and ~Madhukrishna Chakraborty ${ }^{2}$ \footnote{\url{ chakmadhu1997@gmail.com}}\\
	${ }^{1}$ Department of Mathematics, Brainware University, Kolkata-700125, West Bengal, India\\
	${ }^{1}$ Shinawatra University, Thailand\\
	${ }^{1}$ INTI International University, Malaysia\\
	${ }^{2}$ Department of Mathematics, Techno India University, Kolkata-700091, West Bengal, India}
\date{}
	\maketitle
\begin{abstract}
Traversable wormholes (TWHs) remain one of the most intriguing predictions of General Relativity (GR), offering passage through space-time. However, their existence requires the violation of the null energy condition, making their detection a bit challenging. The essay aims at showing the new avenues probing the observational prospects of TWHs via quasinormal modes and grey body factors, the two fundamental aspects in wave dynamics. The role of QNMs in characterizing the ringdown phase of perturbations and the grey body factors in determining transmission probabilities through wormhole barriers has been investigated. Given their distinct spectral imprints, these features provide a potential means to distinguish wormholes from black holes in gravity wave observations. Advances in high-precision interferometry and multi-messenger astronomy may soon offer crucial insights into the existence of these exotic structures.
\end{abstract}

Keywords : Traversable Wormholes ; Quasinormal modes ; Gravity waves ; Grey Body factor.

 ***\underline{\small Essay received Honorable Mention at the Gravity Research Foundation 2025 Awards for Essays on Gravitation}
\newpage
Wormhole (WH) is a hypothetical entity in astrophysics, originally formulated in Einstein gravity and is considered topologically as a shortcut passage between two distant universes or two distance space-time regions of the same Universe \cite{Flamm1916}. The existence of a traversable wormhole (TWH) requires matter violating the Null Energy Condition (NEC), particularly around the throat \cite{EinsteinRosen1935}, \cite{Ellis1973}. Immediately, after the formulation of Einstein gravity Flamm \cite{Flamm1916} gave a solution to the field equations-the first ever theoretical prediction of WH. Later Einstein and Rosen \cite{EinsteinRosenBridge} gave another solution similar to bridge like nature and is commonly known as Einstein-Rosen bridge. The idea of traversability introduced by Morris-Thorne is very popular due to simplified and physically transparent nature \cite{MorrisThorne1988} and argued that traversability can be ensured by exotic matter surrounding the throat of the WH. As a time-like or null test particle can pass through the TWH in finite time so one may observe their response in the form of the wave scattering and quasinormal ringing.

\begin{figure}[h!]
	\centering\includegraphics[height=6.5cm,width=7.5cm]{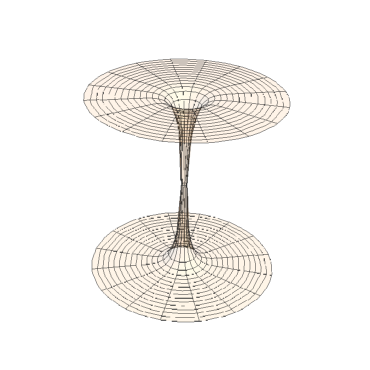}
	\caption{Diagrammatic representation of a Wormhole
	}\label{f1}
\end{figure}
The static spherically symmetric Lorentzian traversable WH (LTWH) introduced by Morris-Thorne is characterized by its shape function $b(r)$ and red-shift function $\Phi(r)$. An extension to axially symmetric TWH is introduced by Teo \cite{Teo1998}. Being an astrophysical entity, the interesting feature of WH is the light deflection character both in strong as well as in the weak deflection limit for a WH space-time \cite{Perlick2015}. Interestingly, WHs behave like BHs in the context of photon trajectories around them. It is well known that around the BHs, the photons can either fall into the BH or scattered away from it to the infinity. In between there are critical photon orbits which separate the above two sets and are termed as unstable photon orbits. They appear to a distant observer as shadow \cite{Nedkova2013}. The situation
should be similar in case of WHs and one should have WH shadow. In the context of BH , there is a strong observational evidence regarding shadows. This is because, the shadows of $M 87 *$ and $\operatorname{SgrA*}$ have been found by Event Horizon Telescope (EHT) \cite{Afrin2023}. As a result, the astrophysicists are very excited to have an observational evidence for other compact objects including WH. From perturbative point of view, the WHs are gravitationally stable, provided TWH is supported by some exotic matter (dark energy). It is speculated that in course of inflationary scenario the primordial microscopic WHs evolve to macroscopic size. Also, assuming existence of WHs one may observe new stars considering WH as a bubble. In analogy to supermassive BHs, accretion disk should be visible across a rotating WH surrounded by some luminous matter. Hence, the detection of WHs may be possible through indirect evidences like gravitational lensing, unique signatures in quasinormal modes and/ or anomalous energy distributions inconsistent with Black Hole (BH) models. The century waited theoretical prediction of GR has been observationally verified by LIGO and VIRGO collaborators \cite{Abbott2017}, \cite{Abbott2016}. It is well known today that in gravity waves (GW) produced during the collision of BHs, the final stage of the ringdown phase is characterized by the quasinormal modes \cite{Konoplya2025}. The study of these QNMs reveal that it is possible to have such QNMs due to the existence of WHs also \cite{Konoplya2025}. The results of the detection of GW not only prove the existence of BH but also hints a distinction between a BH and WH or a possible collision between a BH and a WH \cite{Konoplya2025}. Interestingly, since the discovery of GW, astrophysicists have been trying to associate the observational aspects with the theoretical predictions of compact objects. Specifically, the real part of the QNMs are related to the angular velocity of the last circular null geodesics and a correspondence between the QNMs and the strong lensing limit. Subsequently, it has been shown that the shadow radius is related to the real part of QNMs not only for static/rotating BHs but also is true for static/ rotating WHs - a nice interplay between GW astronomy and the shadow of compact objects. More specifically, for a distant observer the GW from a compact object behaves as massless scalar field propagating along the last unstable null orbit and gradually approaching to spatial infinity.

Line element of a general spherically symmetric static WH space-time of the Morris Thorne class (traversable) is given by \cite{Mehdizadeh2015}-\cite{Halder2019}
\begin{equation*}
d s^{2}=-e^{2 \Phi(r)} d t^{2}+\frac{1}{\left(1-\frac{b(r)}{r}\right)} d r^{2}+r^{2} d \Omega_{2}^{2} \tag{1}
\end{equation*}
where $d \Omega_{2}^{2}=d \theta^{2}+\sin ^{2} \theta d \phi^{2}, \Phi(r)$ is called the red-shift function and it is a function of the radial coordinate $r$ such that $r_{0} \leq r<\infty . r_{0}$ is the radius of the WH throat. Red-shift function is always finite everywhere to avoid horizon or singularity. The function $\Phi(r)$ is used to detect the red-shift of the signal by a distant observer and gives information about the radial tidal force. $b(r)$ is called the\\
shape function that determines the shape of the WH. There are certain restrictions on $b(r)$. They are listed as follows:
  \begin{enumerate}
    \item $b\left(r_{0}\right)=r_{0}$ is the throat condition and $b(r)<r$ for $r>r_{0}$ (metric condition).
    \item Flairing out condition: $b(r)-r b^{\prime}(r)>0$ where $b^{\prime}(r)=\frac{d b(r)}{d r}$.
    \item Asymptotic flatness: $\frac{b(r)}{r} \rightarrow 0$ as $r \rightarrow \infty$.
  \end{enumerate}
The ECs that play a crucial role in formation of a feasible WH are given by \cite{{Chakraborty2024EPJC}}

  \begin{enumerate}
    \item Null Energy Condition (NEC): $\rho+p_{r} \geq 0, \rho+p_{t} \geq 0$.
    \item Weak Energy Condition (WEC): $\rho \geq 0, \rho+p_{r} \geq 0, \rho+p_{t} \geq 0$.
    \item Strong Energy Condition (SEC): $\rho+p_{r} \geq 0, \rho+p_{t} \geq 0, \rho+p_{r}+2 p_{t} \geq 0$.
    \item Dominant Energy Condition (DEC): $\rho-\left|p_{r}\right| \geq 0, \rho-\left|p_{t}\right| \geq 0$.
  \end{enumerate}
Actually, the constraint of minimum radius at the throat with traversability criteria impose huge tension at the throat that is further balanced by the matter violating NEC at the throat i.e, $\rho+p_{r}<0$ i.e, $b(r)-r b^{\prime}(r)>0$ which is essentially the criteria for WH traversability. Raychaudhuri equation (RE) plays a very important role in understanding traversability of WHs and their connection to the violation of ECs in GR. RE describes evolution of a congruence of null geodesics and is given by \cite{Chakraborty2024IJGMMP}

\begin{equation*}
\frac{d \Theta}{d \lambda}=-\frac{\Theta^{2}}{2}-2 \sigma^{2}+2 w^{2}-R_{\mu \nu} u^{\mu} u^{\nu} \tag{2}
\end{equation*}

where $\Theta$ is called the expansion scalar, $\sigma$ is called the anisotropy scalar, $w$ is the vorticity scalar and $\lambda$ is the affine parameter. Focusing theorem (FT) is the most important consequence of RE and it states that if $R_{\mu \nu} u^{\mu} u^{\nu} \geq 0$, then the congruence of null geodesics focus within some finite value of the affine parameter. Further, Landau \cite{KarSenGupta2007} showed in his book that defocusing avoids singularity. Now, $R_{\mu \nu} u^{\mu} u^{\nu}<0$ is the criteria for defocusing and it is the condition for violation of ECs that leads to traversability in WHs. Therefore, violation of ECs, defocusing of geodesic, no horizon/ singularity and traversability are all equivalent. For details regarding RE and geodesic FT one may refer to \cite{KarSenGupta2007}.

From mathematical point of view, the detection of Lorentzian TWH may be possible due to the passage of test fields in the neighbourhood of the WH by observing their response as waves scattering and quasinormal ringing. Further, advanced observational mechanisms namely high resolution interferometeres, precise timing of astrophysical signals and multi-messenger astronomy may identify such\\
signatures. Note that the existence of exotic matter as well as the stringent constraint due to stability analysis put more challenges to observational aspect of detecting WHs. It is speculated that the detection of WH may revolutionize our idea about space-time topology, cosmic connectivity leading to an inside of quantum gravity as well as the fundamental nature of the Universe.

Suppose among the distant space-time regions which are connected by a traversable WH , one space-time region is illuminated by a light source while there is no light source near the throat of the WH in the space-time region on the other side. Usually, one may have two types of orbits in the first region namely \cite{Chakraborty2025DarkUniv}, \cite{Bugaev2021}\\
(a) photons enter the WH and pass through its throat\\
(b) photons scatter away from the WH to infinity.

Thus, a distant observer in the first region will only be able to see the scattered photons while photons captured by the WH will appear to him as a dark spot. This dark spot/ part in the luminous background is termed as the shadow of the WH. Mathematically, for scattering of the photons the radial motion should have a turning point characterized by $\frac{d r}{d \lambda}=0$ i.e, $V_{e f f}=0$. Further, the critical orbit between the scattered and plunged orbits is characterized by the maximum of the effective potential. This critical orbit is a spherical orbit and it is unstable in nature as due to a small perturbation it may either be an escape or captured orbit. Thus, the critical orbit is characterized by $V_{e f f}=0=\frac{d V_{e f f}}{d r}$ and $\frac{d^{2} V_{e f f}}{d r^{2}} \leq 0$. So for WH configuration, the radius of the photon sphere $r_{p h}$ which locates the apparent image of the photon rings satisfies \cite{Bugaev2021}-\cite{Neto2023}
\begin{equation*}
r \Phi^{\prime}(r)=1 \tag{3}
\end{equation*}
i.e,
\begin{equation*}
e^{\Phi(r)}=\Phi_{0} r \tag{4}
\end{equation*}
In fact, $r_{p h}$ is the largest real root of the equation $r \phi^{\prime}(r)=1$ and it identifies the shadow radius as
\begin{equation*}
r_{s h}=\left.r e^{-\Phi(r)}\right|_{r=r_{p h}} \tag{5}
\end{equation*}
where $r_{s h}$ is the radius of the shadow. For convenience, we assume that the observer is situated far away from the WH so that WH shadow radius can be expressed as \cite{Bugaev2021}

\begin{equation*}
r_{s h}=r_{p h} e^{-\Phi\left(r_{p h}\right)}=\sqrt{X^{2}+Y^{2}} \tag{6}
\end{equation*}

where in the observer's frame ( $X, Y$ ) are identified as the celestial coordinates. Now, suppose $r_{0}$ is the distance of the observer from the WH and $\theta_{0}$ is the observer's angular coordinate (or inclination angle) i.e, ( $r_{0}, \theta_{0}$ ) is the location of the observer, then ( $X, Y$ ) can identify the boundary curves of\\
the WH shadow (i.e, the apparent shape of the shadow) and are related to the observer's coordinates $\left(r_{0}, \theta_{0}\right)$ as \cite{Bugaev2021}-\cite{Neto2023}
\[
\begin{array}{r}
X=\lim _{r_{0} \rightarrow \infty}\left(-r_{0}^{2} \sin \theta_{0}\right) \frac{d \phi}{d r} \\
Y=\lim _{r_{0} \rightarrow \infty}\left(r_{0}^{2} \frac{d \theta}{d r}\right) \tag{8}
\end{array}
\]
i.e,
\[
\begin{array}{r}
X=-\frac{\mu}{\sin \theta_{0}}  \tag{9}\\
Y=\sqrt{\nu-\mu^{2} / \sin ^{2} \theta_{0}}
\end{array}
\]
where (expressions of the four velocities have been used) $\lim _{r_{0} \rightarrow \infty} e^{-2 \Phi\left(r_{0}\right)}=$ constant is considered and $\mu, \nu$ are called impact parameters \cite{Bugaev2021}-\cite{Neto2023}. From the above interrelations among the celestial coordinates and the impact parameters one may form the shadow of the traversable WH. For a diagrammatic representation of the shadow one has to plot $X$ vs $Y$ to identify the shadow's boundary (for simplicity one may choose the equatorial plane $\theta_{0}=\frac{\pi}{2}$ ).

Quasinormal modes (QNMs) and grey body factors are very crucial concepts in the arena of wave dynamics in curved space-time. They describe how perturbations such as Gravitational waves (GWs) behave near a compact astrophysical object. QNMs are the characteristic oscillations of a BH or other compact objects that result from perturbations. Grey body factors describe how incoming radiation from BH is partially transmitted and partially reflected due to the curvature of space-time. The correspondence between QNMs and grey body factors has been established for spherically symmetric asymptotically flat or de-Sitter BHs \cite{Bolokhov2024}. Grey body factors might exhibit greater stability against deformations of the near horizon geometry \cite{Konoplya2010} as compared to the overtones of QNMs which are highly sensitive to such deformations. The correspondence between QNMs and grey body factors for BHs have been found using WKB approach. Given that, the boundary conditions for QNMs and grey body factors are the same for BHs and WHs, it is reasonable to consider such similar correspondence in case of WHs too. In fact, in terms of tortoise coordinate $r_{T}$, we can distinguish WH from a BH. $-\infty<r_{T}<+\infty$ in case of BH represents the event horizon and the one side of the asymptotic region, while $-\infty<r_{T}<+\infty$ takes into account the two asymptotically flat space-time regions of a WH with no horizon. Though, the literature on QNMs of WHs is quite vivid, the calculation of grey body factors is case-specific.

In the context of gravitational waves (GW) astronomy, the study of quasinormal modes (QNMs) and grey body factors for WHs has enhanced interest in recent times \cite{Bolokhov2024}, \cite{Konoplya2010}. These quantities identify the response of the space-time to perturbations and its observational characteristics. The damped oscillations of space-time are represented by QNMs and are uniquely characterized by the geometry of the WH and boundary conditions. On the other hand, grey body factors measure the transmission probability of waves propagating through the WH's effective potential barrier. Thus, it is speculated that both the QNMs and grey body factors may be considered as tools to distinguish WHs from BHs in the context of GW astronomy and related observational studies. One may perturb a WH geometry given by (1), considering small deviations in the background space-time geometry or equivalently by examining the evolution of test fields in the underlying geometry. As a result, the perturbation field follows Schrödinger-like wave equation as \cite{Bolokhov2024}, \cite{Konoplya2010}
\begin{equation*}
\left(\frac{d^{2}}{d r_{T}^{2}}+\omega^{2}-V\left(r_{T}\right)\right) \Psi\left(r_{T}\right)=0 \tag{11}
\end{equation*}
Here, $\omega$ stands for the frequency of oscillation, $V$ is the effective potential that depends on the WH geometry and the tortoise coordinate $r_{T}$ has the expression
\begin{equation*}
r_{T}(r)=\int \frac{d r}{e^{\Phi} \sqrt{1-\frac{b(r)}{r}}} \tag{12}
\end{equation*}
The explicit form of the effective potential is given by
\begin{equation*}
V(r)=e^{2 \Phi}\left(\frac{l(l+1)}{r^{2}}-\frac{\left(r b^{\prime}-b\right)}{2 r^{3}}+\frac{\Phi^{\prime}(r)}{r}\left(1-\frac{b(r)}{r}\right)\right) \tag{13}
\end{equation*}
for massless scalar field and
\begin{equation*}
V(r)=e^{2 \Phi} \frac{l(l+1)}{r^{2}} \tag{14}
\end{equation*}
for electromagnetic field. Here, $l$ (the angular momentum quantum no.) represents the multipole number.
The QNMs $(\operatorname{Re}(\omega)+i \operatorname{Im}(\omega))$ of WHs are the solutions of the above wave equation (11) with the purely outgoing boundary conditions at infinity i.e,
\begin{equation*}
\Psi\left(r_{T}\right) \approx \Psi_{I} e^{ \pm i \omega r_{T}}, r_{T} \rightarrow \pm \infty \tag{15}
\end{equation*}
It is to be noted that, the QNM gives information about the evolution of the perturbed field as follows: the real part of $\omega$ identifies the oscillation of the signal while the imaginary part measures the damping factor for the energy loss of gravitational radiation. More specifically, if $\Psi\left(r_{T}\right) \approx \Psi_{I} e^{i \omega r_{T}}, r_{T} \rightarrow \pm \infty$ then $\operatorname{Im}(\omega)<0$ gives unstable perturbations with exponential growth of the perturbed field and $\operatorname{Im}(\omega)>0$, implies stability of the perturbed field. Reversely, if $\Psi\left(r_{T}\right) \approx \Psi_{I} e^{-i \omega r_{T}}, r_{T} \rightarrow \pm \infty$ then $\operatorname{Im}(\omega)<0$ gives stable perturbations and $\operatorname{Im}(\omega)>0$ gives unstable perturbations.
On the other hand, due to Schrödinger like nature of wave equation (11) the WKB approximation is preferable in this arena.
Thus, using WKB approach the QNM frequency $\omega$ near the WH throat takes the form
\begin{equation*}
\omega=\frac{e^{\Phi\left(r_{0}\right)}}{r_{0}}\left(l+\frac{1}{2}\right)-i\left(n+\frac{1}{2}\right) \frac{e^{\Phi\left(r_{0}\right)}}{\sqrt{2} r_{0}}+O\left(l^{-1}\right) \tag{16}
\end{equation*}
Hence the real part has a relation with the shadow radius as
\begin{equation*}
R e(\omega)=\frac{1}{r_{s h}}\left(l+\frac{1}{2}\right) \tag{17}
\end{equation*}
The above expression shows that the shadow radius depends not only on the WH throat but also on the outer light ring $r_{p h}>r_{0}$. The above interrelation is exact in the eikonal limit (large $l$ ), while the above relation is very close for small values of $l$ as well. Also, as in the eikonal limit the electromagnetic and the scalar field behave identically so the interrelation between shadow radius and the real part of the QNMs remains same for scalar field propagation also. Now, due to smoothness of $b(r)$ and $\Phi(r)$, it is possible to have the Taylor series expansion about the WH throat at $r=r_{0}$ as
\begin{gather*}
b(r)=b_{0}+b_{1}\left(r-r_{0}\right)+b_{2}\left(r-r_{0}\right)^{2}+\ldots  \tag{18}\\
\Phi(r)=\Phi_{0}+\Phi_{1}\left(r-r_{0}\right)+\Phi_{2}\left(r-r_{0}\right)^{2}+\ldots \tag{19}
\end{gather*}
It is to be noted that, due to asymptotic flatness it may be desirable to have a pre-factor for correct behavior. However, in the present context as we are studying with respect to the WH throat so pre-factor is not required. Now, for large $l$ one may have analytic form for the QNMs using WKB formula. The formula for the QNM frequencies to the $K-t h$ order in perturbation takes the form
\begin{equation*}
i \frac{\omega^{2}-V_{m}}{\sqrt{-2 V_{m}^{\prime \prime}}}-\sum_{i=2}^{K} \Gamma_{i}=n+\frac{1}{2} \tag{20}
\end{equation*}
Here, $V_{m}$ is the maximum of the potential, its second order derivative w.r.t the tortoise coordinate is denoted by $V_{m}^{\prime \prime}$, all higher order corrections are contained in $\Gamma_{i}^{\prime} \mathrm{s}$ and $K$ stands for the order of the WKB approximation. It is important to note that larger $n$ does not give a better approximation for the quasinormal frequencies. In fact, the value of $K$ for best value of quasinormal frequency is not unique it depends on $(n, l)$. Further, one can associate the shape of WH with the imaginary part of QNM as $\operatorname{Im}(\omega)=\frac{\sqrt{\left(b_{1}-1\right)\left(b_{0} \Phi_{1}-1\right)}}{\sqrt{2} r_{s h}}$. Thus, the imaginary part which governs the damping or decay of the wave amplitude over time is also characterized by the shadow radius. To be precise, larger shadow radius indicates less damping. Also, $b^{\prime}\left(r=b_{0}\right)=1$ gives $\operatorname{Im}(\omega)=0$ i.e, we have pure real modes as the standing waves of an oscillating string with fixed ends at the throat.

On the other hand, to examine the wave scattering processes in the vicinity of the WH, one has to take into account the interaction of waves with potential barrier surrounding the WH (compact\\
object). Consequently, waves are partially reflected off the barrier and partially transmitted through it. The grey body factors take care of both these events and it does not depend on whether the wave originates in the asymptotic region in the other universe or it arrives from spatial infinity. Thus, here the boundary conditions are modified as

\[
\begin{array}{r}
\Psi=e^{-i \Omega r_{T}}+R e^{i \Omega_{r}}, r_{T} \rightarrow+\infty \\
\Psi=T e^{-i \Omega r_{T}}, r_{T} \rightarrow-\infty \tag{21}
\end{array}
\]

with $R$ and $T$ as reflection and transmission coefficients. Here, the frequency $\Omega$ is real and continuous for scattering phenomena and is distinct from the complex and discrete quasi normal mode frequency. The transmission coefficient $T$ is usually termed as the grey body factor as it measures the fraction of the wave which traverses the potential barrier and as a result it contributes to the emission of radiation from the compact objects (BH or WH). Thus, one may write the grey body factor corresponding to the angular momentum number $l$ as

\begin{equation*}
\Gamma_{l}(\Omega)=|T|^{2}=1-|R|^{2} \tag{22}
\end{equation*}

Using WKB expression for the grey body factor one has

\begin{equation*}
\Gamma_{l}(\Omega)=\frac{1}{1+e^{2 \pi i k}} \tag{23}
\end{equation*}

Now, the effective potential can be expanded in powers of $l$ as

\begin{equation*}
V\left(r_{T}\right)=l^{2} V_{0}\left(r_{T}\right)+l V_{1}\left(r_{T}\right)+l^{-1} V_{2}\left(r_{T}\right)+\ldots \tag{24}
\end{equation*}

and as a result the eikonal approximation can be derived from the first order WKB approximation as

\begin{equation*}
\Omega=l \sqrt{V_{00}}-i k \sqrt{\frac{-V_{00}^{\prime \prime}}{2 V_{00}}}+O\left(l^{-1}\right) \tag{25}
\end{equation*}

where $V_{00}$ is the value of the function $V_{0}\left(r_{T}\right)$ in its maximum and $V_{00}^{\prime \prime}$ is value of second derivative of $V_{0}\left(r_{T}\right)$. As a consequence, $k$ can be expressed as a function of the real frequency $\Omega$

\begin{equation*}
-i k=\frac{\Omega^{2}-l^{2} V_{00}}{l \sqrt{-2 V_{00}}}+O\left(l^{-1}\right)=-\frac{\Omega^{2}-\operatorname{Re}\left(\omega_{0}\right)^{2}}{4 \operatorname{Re}\left(\omega_{0}\right) \operatorname{Im}\left(\omega_{0}\right)}+O\left(l^{-1}\right) \tag{26}
\end{equation*}
Hence, one may associate the transmission coefficient to the grey body factors $\Gamma_{l}(\Omega)$ with the fundamental mode $\omega_{0}$ as
\begin{equation*}
\Gamma_{l}(\Omega)=|T|^{2}=\left(1+e^{2 \pi \frac{\Omega^{2}-R e^{2} \omega_{0}}{4 R e\left(\omega_{0}\right) I m\left(\omega_{0}\right)}}\right) \tag{27}
\end{equation*}
It is to be noted that the above relation is exact in the eikonal limit $l \rightarrow \pm \infty$ and an approximate one for small ' $l$ '. To find the interrelation between shadow radius with grey body factor we consider equations (18)-(26) to get
\begin{equation*}
V_{00}=e^{2 \Phi_{0}}\left(\frac{4 l(l+1)}{2 r_{0}^{4}}\left(\left(2 r_{0}^{2}\left(\Phi_{2}+\Phi_{1}^{2}\right)\right)-1\right)+\frac{b_{2}}{r_{0}^{3}}\left(1-2 \Phi_{1} r_{0}\right)-\frac{3 b_{3}}{r_{0}^{2}}\right) \tag{28}
\end{equation*}
Choosing $\Phi_{2}+\Phi_{1}^{2}=\frac{1}{2 r_{0}^{2}}$ and $\frac{3 b_{3}}{r_{0}^{2}}+\frac{b_{2}}{r_{0}^{3}}=0$, we get
\begin{equation*}
V_{00}^{\prime \prime}=-l(l+1) b_{2}^{2}\left(\frac{e^{\Phi_{0}}}{r_{0}}\right)^{2} \tag{29}
\end{equation*}
Thus,
\[
\begin{array}{r}
\operatorname{Re}\left(\omega_{0}\right)=l\left(\frac{e^{\Phi_{0}}}{r_{0}}\right) \sqrt{l(l+1)} \\
\operatorname{Im}\left(\omega_{0}\right)=\frac{b_{2}}{2 \sqrt{2}} \\
\Gamma_{l}(\Omega)=1+e^{\Lambda} \tag{32}
\end{array}
\]
where,
\begin{equation*}
\Lambda=\frac{\pi}{\sqrt{2}}\left(\frac{\Omega^{2}-\frac{l^{3}(l+1)}{r_{0}^{2}} e^{2 \Phi_{0}}}{b_{2} \frac{e^{\Phi_{0}}}{r_{0}} \sqrt{l^{3}(l+1)}}\right) \tag{33}
\end{equation*}

Equivalently, the above expression in terms of shadow radius can be written as

\begin{equation*}
\Lambda=\frac{\pi}{\sqrt{2}}\left(\frac{\Omega^{2}-\frac{l^{3}(l+1)}{r_{s h}^{2}}}{b_{2} \frac{1}{r_{s h}} \sqrt{l^{3}(l+1)}}\right) \tag{34}
\end{equation*}
From the expression of grey body factor in (32), it is clear that, as the damping increases (reversely the radius of the shadow decreases), at some value of shadow radius we will have $\Omega^{2}<\frac{l^{3}(l+1)}{r_{s h}^{2}}$ making $\Lambda<0$. In this way as $\Lambda \rightarrow-\infty, \Gamma_{l} \rightarrow 1$. This may be interpreted as: "Lesser is the radius of the shadow, more is the effect of damping and grey body factor tends to be unity".

To conclude, the study of QNMs and grey body factors offers a promising avenue in identifying TWHs in astrophysical observations. The relationship between shadows and quasinormal modes (QNMs) of a compact object, particularly wormholes, can be established in the way these objects interact with light and perturbations in the surrounding space-time. Quasinormal modes and greybody factors depend on how perturbations propagate along geodesics in the curved wormhole space-time. While black holes and wormholes exhibit similar perturbation responses, subtle differences in their QNM spectra and grey body factors could serve as key discriminators. Wormholes can mimic black hole shadows but have different quasinormal mode (QNM) spectra due to the absence of a true horizon. This is the reason why the study of QNMs near a compact object can enable us to distinguish them. Furthermore, there is a clear analogy between QNMs and the shadow of a wormhole. The shadow size is inversely related to the QNM frequency $\Omega$, directly linking to the shadow’s boundary. Moreover, the imaginary part of the QNM is also inversely related to the shadow radius, connecting it with damping effects. The dominant QNM frequencies are closely tied to the photon sphere’s properties.
In wormholes, the effective photon potential determines both the shadow boundary and the ringdown frequencies of perturbations, thereby linking these two key observational signatures. Additionally, the grey body factor tends to unity when the shadow radius is very small, while $\Gamma_{l}>>1$ for large shadow radius. Thus, the essay showcases a unification of these two observational features in explaining a feasible WH configuration.
Theoretical progress in stability analysis, coupled with advancements in gravitational wave astronomy, enhances the prospects for observationally distinguishing wormholes from other compact objects. If detected, such structures would not only revolutionize our understanding of space-time topology but also provide deeper insights into quantum gravity and the fundamental nature of the universe—or more precisely, they would enable us to understand the dynamics of the universe and above all answer the question we started with: “Wormholes: Myth or Reality?”
\section*{Acknowledgment}
S.C. thanks the Department of Mathematics, Brainware University, West Bengal, and M.C. thanks Techno India University, West Bengal, for providing research facilities. S.C. also thanks Shinawatra University, Thailand, and INTI International University, Malaysia, for the research fellowship program.

\end{document}